\pdfoutput=1
\documentclass[aip,apl,reprint]{revtex4-1}
\usepackage{graphicx}
\usepackage{dcolumn}
\usepackage{bm}
\usepackage{epstopdf}
\usepackage{color}
\draft 
\begin{document}
\title{Single photon emission from droplet epitaxial quantum dots 
in the standard telecom window around a wavelength of 1.55~$\mu$m} 
%
%
%
\author{Neul~Ha}
\email{ha.neul@nims.go.jp}
\author{Takaaki Mano}
\author{Samuel Dubos}
\author{Takashi Kuroda}
\email
{kuroda.takashi@nims.go.jp}
\author{Yoshiki Sakuma}
\author{Kazuaki Sakoda}
\affiliation{National Institute for Materials Science, 1-1 Namiki, Tsukuba 305-0044, Japan}


\date{\today}

\begin{abstract}
We study the luminescence dynamics of telecom wavelength InAs quantum dots grown on InP(111)A by droplet epitaxy. The use of the ternary alloy InAlGaAs as a barrier material leads to photon emission in the 1.55~$\mu$m telecom C-band. The luminescence decay is well described in terms of the theoretical interband transition strength without the impact of nonradiative recombination. The intensity autocorrelation function shows clear anti-bunching photon statistics. The results suggest that our quantum dots are useful for constructing a practical source of single photons and quantum entangled photon pairs. 
\end{abstract}

\maketitle 

\noindent\textcolor{blue}{%
This is the version of the article before peer review, as submitted to \textit{Applied Physics Express}. The final published version will be available as an Open Access article from the journal's site.}

A source of single photons and quantum entangled photon pairs is a key device in vast quantum technologies. Semiconductor quantum dots are expected to serve as photon sources that can be operated very efficiently and deterministically. Numerous efforts have already been made to develop a practical quantum dot photon source. However, photon emission in the standard telecom band, particularly around a wavelength of 1.55~$\mu$m, which is the maximum transmission window of silica optical fibers, is a material challenge. 
Careful growth optimization is required to achieve a 1.55~$\mu$m emission
\cite{Takemoto_JAP07,Takemoto_SciRep15,Miyazawa_APL16,Birowosuto_SciRep12,SkibaSzymanska_PhysRevAppl17,Muller_NatCom18,Benyoucef_APL13,Paul_APL17,Dusanowski_APL14,Dusanowski_APL16}. Nevertheless, the well-known quantum dot growth based on the Stranski-Krastanow mode leads to an asymmetric dot shape, which is not favorable for entangled pair generation. 

The problem is ideally solved by using droplet epitaxy, which offers considerable freedom regarding the choice of materials and substrates \cite{Gurioli_NatMat19}. The application of a $C_{3v}$ symmetric (111)A surface to the growth substrate results in the creation of almost perfectly symmetric quantum dots, which can work in both the visible wavelength region \cite{Mano_APEx10,Ha_APL19} and the infrared telecom wavelength region \cite{Ha_APL14,Liu_PRB14}. Recently, the emission wavelength has been extended beyond 1.5~$\mu$m for InAs dots embedded in InAlGaAs on InP(111)A \cite{Ha_APEx16}. However, the previous samples were not sufficiently optimized: the dot density was too high for a single quantum dot to be isolated using standard micro optics. Moreover, the dot size distribution is relatively large so that careful dot selection is required to find a dot that emits at 1.55~$\mu$m. 

Here, we extend the droplet epitaxy scheme to achieve a purely 1.55~$\mu$m photon emission. We introduce the high temperature crystallization protocol, which has recently been applied to the GaAs material system \cite{Jo_CDG12,BassoBasset_Nano17}, to the InAs/InP material system in order to improve the dot morphology property. The use of a state-of-the-art superconducting photon detector, together with an efficient dot sample, allows us to investigate single photon emission dynamics in the standard telecom C-band. 

\begin{figure}
\includegraphics[width=7.5cm]{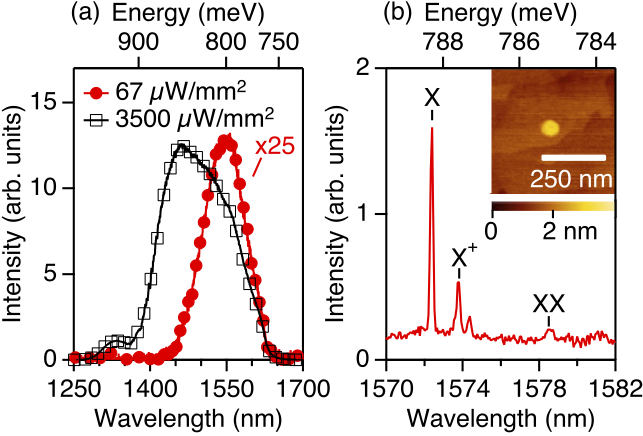}
\caption{\label{fig_afm}(Color online) (b) The luminescence spectra of a large ensemble of InAs quantum dots in In$_{0.52}$Al$_{0.12}$Ga$_{0.36}$As /InP(111)A at 12~K at different excitation powers. (b) The luminescence spectra of a single isolated InAs dot with a cw excitation of 40~nW. We focus on this dot in our time-resolved study. The inset shows an atomic force microscope image of the dot surface.}
\end{figure}

The quantum dot sample is grown on Fe-doped semi-insulating InP(111)A using a solid source molecular beam epitaxy machine. After depositing a lattice-matched In$_{0.52}$Al$_{0.12}$Ga$_{0.36}$As barrier with a thickness of 200~nm at 490~$^{\circ}$C, we grow 0.5 of a monolayer (ML) of InAs at 490~$^{\circ}$C. 
We then supply 0.25~ML of indium at a growth rate of 0.16 ML/s at 400~$^{\circ}$C, which leads to the formation of indium droplets on InAlGaAs(111)A. Next, an As$_4$ flux ($9\times10^{-5}$~Torr) is supplied at 400~$^{\circ}$C to crystallize InAs dots from indium droplets. After \textit{in vacuo} annealing at 450~$^{\circ}$C for 5~min, the dots are capped by In$_{0.52}$Al$_{0.12}$Ga$_{0.36}$As with a thickness of 100~nm. A notable point in this sequence is the crystallization temperature, which we set higher than that of the standard protocol. The small diffusion length of group-III adatoms suppresses the transformation from dots to layers even at 400~$^{\circ}$C, leading to the formation of dots with a high crystalline quality. 

We measure the stationary- and time-resolved responses of photoluminescence from single InAs dots. For the stationary study we use a semiconductor laser diode at a wavelength of 980~nm as a cw excitation source. For the time-resolved study we use a ps mode-locked titanium sapphire laser whose wavelength is tuned to 900~nm as a pulsed source. The laser light is focused on the sample using a microscope objective lens with a numerical aperture of 0.65 (Olympus LCPlan50xIR). The luminescence signal is collected by the same lens, passed through a dichroic beam splitter, and coupled to a single mode optical fiber that has a mode field diameter of 9~$\mu$m at a wavelength of 1.3~$\mu$m. The fiber output is fed into a 50~cm spectrometer that consists of a 600 line/mm grating. 

The luminescence signal is spectrally analyzed using a cooled InGaAs photodiode array (Andor iDus 491) and temporally resolved using a superconducting single-photon detector (Single Quantum Eos) with a fast-response time-to-digital converter (PicoQuant PH300). The polarization state of the input light is adjusted to maximize the detection efficiency. Note that we attach a high-index hemispherical lens ($n=2$) to the sample surface to increase the light collection efficiency \cite{Ha_PRB15}. Thanks to the efficient setup together with the bright sample we achieve a maximum count rate as high as $\sim$50~kHz under saturation conditions. The sample is cooled using a closed cycle cryostat. All the experiments are performed at 8~K unless otherwise noted. 

The inset in Fig.~\ref{fig_afm}(b) shows an atomic force microscope image of the quantum dot surface. It reveals the formation of nearly circular dots without significant elongation. The dot isotropy arises due to the use of the $C_{3v}$ symmetric \{111\} surface as a growth substrate. The quantum dots have a disk-like shape with a diameter of $48\pm 8$~nm and a height of $1.9 \pm 0.3$~nm. These shape parameters are similar to those of droplet epitaxy GaAs dots on AlGaAs(111)A \cite{Mano_APEx10}. The dot density is $\sim 6\times10^8$~cm$^{-2}$ thus making it easy to isolate a single dot without any post-growth processing such as the fabrication of small mesas or apertures. 

Figure~\ref{fig_afm}(a) shows the photoluminescence spectra of the quantum dot ensemble. They were measured using standard long focus optics. For low excitation, the spectrum has a Gaussian-like single peak centered at a wavelength of 1,550~nm. Its full width at half maximum is $\sim$100~nm, which  is more than two times smaller than that of our previous sample targeting a 1.55~$\mu$m emission \cite{Ha_APEx16}. 
Hence, the majority of the dots in the present sample can emit in the telecom C-band. For high excitation, the spectrum shows another broad band that originates from the excited states, as well as an additional peak at 1,350~nm due to carrier recombination in the barrier layer. 

Figure~\ref{fig_afm}(b) shows a typical luminescence spectrum for a single isolated dot. The observed split lines are attributed to neutral excitons (X, 1,572.4~nm), positively charged excitons (X$^+$, 1,573.8~nm), and neutral biexcitons (XX, 1,578.6~nm). The spectral assignment was based on the large number statistics of the multiexciton binding energies in InAs/InAlAs droplet dots \cite{Liu_PRB14}. Note that the present sample frequently shows X$^+$, but rarely shows a negatively charged line. This implies that our sample is slightly \textit{p}-doped possibly due to the residual presence of carbon impurities. 

\begin{figure}
\includegraphics[width=5.5cm]{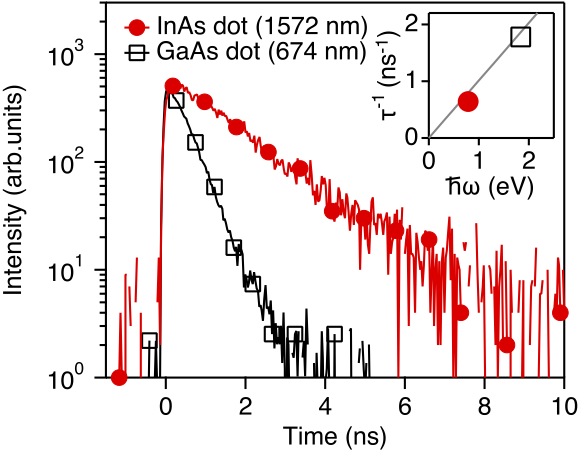}
\caption{\label{fig_decay}(Color online) The luminescence decay of the neutral exciton line of an InAs quantum dot emitting at a wavelength of 1,572~nm with an excitation power of 4~nW (red circles). 
The black square line is the luminescence decay of the neutral exciton line of a GaAs quantum dot embedded in Al$_{0.3}$Ga$_{0.7}$As(111)A, with an emission wavelength of 674~nm. 
The inset shows the decay rate dependence on the emission frequency for the measured GaAs and InAs dots, and the theoretical prediction of Eq.~(1) for $n=3.5$ and $2p^2/m_0 = 20$~eV. }
\end{figure}

Figure~\ref{fig_decay} shows the luminescence decay signal of the X line shown in Fig.~\ref{fig_afm}(b) after short pulsed excitation. The decay signal of a GaAs quantum dot embedded in Al$_{0.3}$Ga$_{0.7}$As(111)A, which we studied previously \cite{Kuroda_PRB13}, is also shown for comparison. Both decay curves are well approximated by straight lines in the semilogarithmic plot, which implies that they follow single exponent functions. The decay time constant of the InAs dot is estimated to be 1.56~ns, which is significantly longer than that of the GaAs dot (0.56~ns). The large difference in the luminescence decay times arises due to the frequency dispersion of the photonic density of states, as discussed below. 

The spontaneous emission rate for atomic transitions (Einstein's A coefficient) is expressed as \cite{Loudon}
\begin{equation}
\tau^{-1} = %
\left( \frac{\mu}{\mu_0}n \right)%
\frac%
{e^2\omega \left\vert \bar{p} \right \vert ^2}%
{3\pi\epsilon_0 \hbar m_0^2 c^3} \, ,
\end{equation}
where $\epsilon_0$ and $\epsilon$ ($\mu_0$ and $\mu$) are the vacuum and relative permittivities (permeabilities), respectively, 
$n$ is the refractive index given by $\sqrt{\epsilon\mu/\epsilon_0\mu_0}$, $\omega$ is the angular frequency of emitted light, and $p$ is the matrix element of the momentum operator $\widehat{\mathbf{p}}=-i\hbar \mathbf{\nabla}$. The above formula was deduced by including the interaction Hamiltonian $H_{\mathrm{int}}=(\mathbf{A \cdot \widehat{p}}+\mathbf{\widehat{p}\cdot A})e/2m_0$, where $\mathbf{A}$ is the quantized vector potential, and the three-dimensional photonic density of states $\rho(\omega)=V\omega^2/\pi^2 c^3$, where $V$ is the normalization volume, to the Fermi's golden rule. 

Note that the matrix element $p$ serves as a band mixing source in the $\mathbf{k}\cdot \mathbf{p}$ perturbation theory, and it is more or less constant for most group-IV, III-V, and II-VI semiconductors, with the Kane energy $2p^2/m_0 \approx 20$~eV \cite{YuCardona}. Consequently, the material dependence in Eq.~(1) appears only in the $\omega$-proportional factor if we assume a constant $n$ value. 
The solid line in the inset of Fig.~\ref{fig_decay} is the model dependence of the emission decay rate on the photon frequency, where we assume that $n=3.5$ and $2p^2/m_0 = 20$~eV. The $\omega$-linear dependence agrees with the measured decay rates of telecom wavelength InAs dots and visible wavelength GaAs dots. Thus, the radiative process of our quantum dots is purely described by the atomic description in Eq.~(1) free from the impact of nonradiative recombination. 

\begin{figure}
\includegraphics[width=7cm]{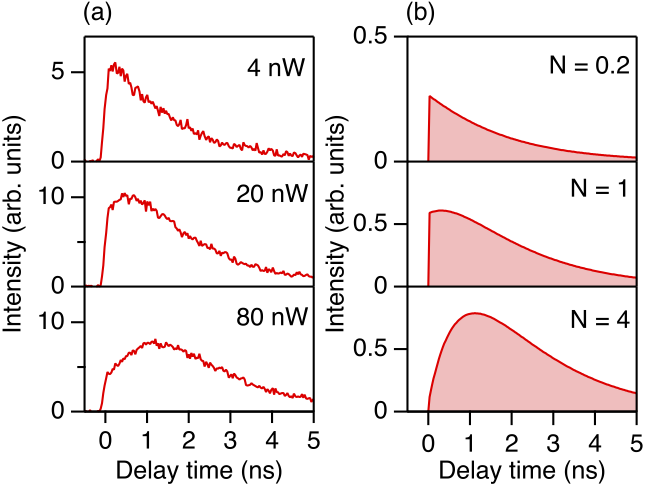}
\caption{\label{fig_cascade} (Color online) (a) Comparison of the transient responses of the luminescence signals of the X line for different excitation powers. (b) Model calculation results for the luminescence transients. $N$ is the average initial exciton number. %
}
\end{figure}
Figure~\ref{fig_cascade}(a) shows the luminescence transients for different excitation powers. The signal observed at 4~nW, i.e., the lowest excitation condition, is identical to that shown by the semilogarithmic plot in Fig.~\ref{fig_decay}. Hence, the decay curve follows a single exponent. When the excitation power is increased to 20~nW, the signal deviates from a monotonic decay, and reveals a significant rise after $t=0$. The rise signature is more evident for 80~nW, where the intensity maximum is substantially delayed by more than 1~ns after excitation. The observed power-dependent evolution arises due to the multiexciton relaxation cascade. We analyze the single exciton spectral line, which is generated only when a single electron and hole pair remains in the dot, following the recombination of all the other pairs. The same luminescence behavior is reported in Refs.~\onlinecite{Dekel_PRB00,Kuroda_PRB02}. Figure~\ref{fig_cascade}(b) shows the numerical simulation results, where we assume a Poissonian distribution for the initial number of excitons \cite{Abbarchi_JAP09}. For simplicity, we deal only with the cascade evolution from XX to X, and we assume that XX decays twice as fast as X, i.e., XX decays like noninteracting two excitons. The simple model reproduces the measured behavior qualitatively. 

\begin{figure}[t]
\includegraphics[width=7cm]{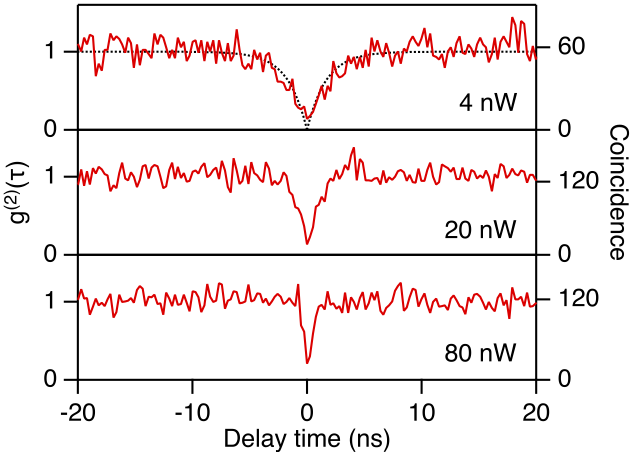}
\caption{\label{fig_cncdnc}(Color online) The intensity autocorrelation function of the X line for different excitation powers. The coincidence number was integrated with a time bin of 256~ps for 6~hours (4~nW), 80~min (20~nW), and 20~min (80~nW). The simulation results are also plotted by the gray broken line. }
\end{figure}

Figure~\ref{fig_cncdnc} shows the intensity autocorrelation function $g^{(2)}(t)$ of the X luminescence line in Fig.~\ref{fig_afm}(b). Here, we adopt the Hanbury Brown and Twiss setup to measure the coincidence of two photons as a function of delay time \cite{Kuroda_APEX08}. The sample is illuminated by cw light. With low excitation at 4~nW, the signal shows a clear antibunching dip, which yields nearly no probability of emitting two photons at the same time. As the delay time is increased the signal recovers from $\sim 0$ to the equilibrium value, i.e., the accidental coincidence number, with a time constant given by the luminescence decay time analyzed in Fig.~\ref{fig_decay} ($\tau = 1.56$~ns). The solid line shows a model $\propto 1-\exp(-\vert t \vert/\tau)$, which agrees with the observed signal. With increasing excitation power, the dip width is observed to decrease, and the signal quickly recovers to the equilibrium level. This is due to the acceleration of the X population recovery for strong excitations \cite{Sasakura_APEX15}. Note that several researchers have reported the emergence of positive bunching correlations superimposed on the antibunching dip \cite{Regelman_PRL01,Kiraz_PRB02,Odashima_JAP17}. However, we do not observe such a signature possibly due to the lower population of neutral XX in our sample. Nevertheless, the value of $g^{(2)}(0)$ is lower than the classical limit of 0.5 at least over the present excitation range, supporting the emission of single photons from this dot. 

In conclusion, we used droplet epitaxy to fabricate InAs quantum dots that emit single photons at wavelengths around 1.55~$\mu$m. Careful growth optimization enabled us to reduce the dot size distribution, and so the majority of the dots could emit photons in the telecom C-band. The exciton lifetime was nearly the same as the theoretically ideal value free from the impact of nonradiative recombination.  The use of a trigonally symmetric InP(111)A substrate led to the formation of nearly circular dots. Thus, our dots can be expected to serve as bright single photon and entangled photon pair sources that will be useful for practical quantum communication applications. 



%
%

%

\begin{acknowledgments}
We acknowledge support from a Grant-in-Aid from the Japan Society for the Promotion of Science (Grant number 16H02203), and Innovative Science and Technology Initiative for Security, ATLA, Japan. 
\end{acknowledgments}


\bibliography{entangleLED.bib,singlePhoton1p55.bib}

\end{document}